\documentclass[11pt]{article}
\usepackage{amsmath,amssymb,graphicx}

\begin{document}

\title{Quantum Phase Transition, Dissipation, and Measurement}
\author{Sudip Chakravarty\\ Department of Physics and Astronomy\\ University of California Los Angeles\\ Los Angeles, California 90095-1547, USA}
\date{\today}
\maketitle
\tableofcontents

\section{Introduction}
This article reflects my  subjective choice of topics. My interest falls into three distinct categories. The intense interest in the problem of dissipation in quantum systems began at the time when macroscopic quantum tunneling and coherence in superconducting quantum interference devices were proposed. This subject has seen considerable developments  over the past three decades both in theory and in experiments.~\cite{Leggett:2002} An important problem in this subject is a particle in a double well potential coupled to a dissipative Ohmic heat bath.~\cite{Leggett:1987} Yet, it is hardly recognized that this problem  constitutes one of the simplest models of quantum criticality.~\cite{Chakravarty:1982} Thus, unsurprisingly, little attention has been paid over the years to experimentally explore this critical phenomenon.  It remains  equally unrecognized  that the same model holds a germ of the quantum to classical transition much discussed in measurement theory.\footnote{I have reluctantly succumbed to the trap of using the vague terminology---measurement theory} Here I have attempted to draw a parallel with 
an even more elementary model of Coleman and Hepp,~\cite{Hepp:1972} which actually  contains no dissipation. Nonetheless the two problems  are united by the dictum that for infinitely many degrees of freedom two distinct states of matter cannot be unitarily related. A particle in a  double well coupled to a dissipative environment is indeed a system with infinite number of degrees of freedom.  Of course, Coleman-Hepp model can be supplemented by dissipation to bring the two models closer, but the basic issue with respect to measurement is clearer in its original form.

 Certain aspects of  quantum phase transitions have recently been shown to be  different from their classical counterparts. A surprise has been  that  Anderson localization of an electron in a random potential and plateau to plateau transition of the integer quantum Hall effect do not fit the traditional framework.~\cite{Yang:1952,Kopp:2007}  Despite field theoretical formulation that mimics a conventional quantum phase transition, there is a glaring issue that is lost in the forest of replica formalism~\cite{Belitz:1994} where the number of replicas have to be analytically  continued to zero. One finds that a cherished criterion that the  ground state energy has to be non-analytic at a quantum phase transition is simply not true for these phenomena~\cite{Edwards:1971,Kopp:2007}.  Instead, one finds that the von Neumann entropy is beautifully non-analytic and in perfect agreement with the scaling theory of localization~\cite{Abrahams:1979} and its generalization to integer quantum Hall systems~\cite{Huckestein:1995}. It is this quantum information theoretical perspective that intrigues me.

Finally, disorder brings some surprises to the subject of quantum phase transition. It can be argued that a whole class of first order quantum phase transitions are rounded by disorder and even converted  into continuous phase transitions~\cite{Goswami:2008}. Mathematical physicists are beginning to pay some serious attention to this problem~\cite{Greenblatt:2009}, which, however, is not a trivial consequence of a similar result for classical first order phase transitions.~\cite{Aizenman:1990} It makes us wonder how many sightings of quantum critical points are actually first order quantum phase transitions in disguise. And, in fact, how should one approach disorder rounded quantum criticality, especially its  dynamics. This is a subject that is in its nascent stage, but is likely to become important in  the near future.

\section{Multiplicity of Dynamical Scales and Entropy}

 A phase transition between two distinct states of matter is characterized by a non-analyticity of the free energy.
This is now the dogma, and rightly so, because there are very few cases where this idea could in principle be challenged.
In the back of our minds we have the picture of  the distribution of zeros of the partition function of the two-dimensional
Ising model. Yang and Lee showed that these zeros lie on a unit circle and in the infinite volume limit they become dense and pinch the real axis at a finite temperature, separating two phases~\cite{Lee:1952}. At zero temperature the ground state energy plays the role of free energy.

This dogma was  scrutinized a great deal   in the context of spin-glass transition~\cite{Binder:1986} where  the interaction between the  spins is a quenched random variable
distributed  according to a probability distribution. But with some tricks involving averaging the free energy (not the partition function) with respect to disorder nothing really changed. Much effort, including those of the  present author~\cite{Singh:1986},  has been spent in describing the spin glass transition as an equilibrium phase transition that we have known to love and cherish.  
When  the real world offers us with a multiplicity of dynamical scales, equilibrium is difficult to attain within a typical observation time scale.  This is nowhere more prominent than in a quantum system where statics and dynamics are intertwined. In fact, the explicit nature of the reservoir to which a system is coupled  can drastically change the nature of a quantum phase transition at zero temperature,\footnote{A quantum critical point is a point at exactly $T=0$, where the correlation length diverges as we tune a coupling constant, resulting
in a non-analyticity in the ground state energy (sharp avoided level crossing)  and/or von Neumann entropy,
as in a transverse field Ising model in one spatial dimension for which there is no finite temperature transition. If, on the other hand, we consider a transition, as for example in a transverse field Ising model in two spatial dimensions, there is a finite temperature transition as well. However, the criticality of the finite temperature transition has the Onsager exponents, because at the critical point the correlation length tends to infinity and the criticality is determined by the classical statistical mechanics of the  two-dimensional Ising model. Because the energy scales tend to zero as the transition is approached, quantum mechanics is unimportant at such long wavelengths. However, there will be a crossover from quantum to classical behavior,  as we approach the criticality at non-zero temperature. So the limit $T \to 0$ can be singular. If we have a first order quantum transition at $T=0$, where an order parameter develops discontinuously, the story is different, because
the correlation length is finite and remains so even if this transition continues to finite temperatures, and in principle quantum mechanics
can play an important role even at non-zero temperatures, if the correlation length is sufficiently short.} and even its very existence can depend upon it.   
 
Before Gibbs, Boltzmann had postulated a 
famous formula that is perhaps not so practical but conceptually important.  It defines
entropy, $S_{B}$, in terms of the available phase space volume, $W$:
\begin{equation}
S_{B} = k_B \ln W.
\end{equation}
where $k_{B}$ is the Boltzmann constant. But how do we find $ W$? It 
appears that in order to find it we must solve the
equations of motion and determine the dynamics of the system. The Boltzmann formula can of course  be reconciled with
the ensemble approach of Gibbs in equilibrium, which is more convenient to use~\cite{Lebowitz:2007}. A  system coupled weakly  to a reservoir can be described by a canonical ensemble when the total system is described by a microcanonical ensemble in an energy shell, for either classical phase space distribution or for quantum density matrices. Surprisingly, a stronger result has been obtained~\cite{Goldstein:2006,Gemmer:2003}, that, in the thermodynamic limit, the reduced density matrix of the system is canonical for the overwhelming majority of  {\em wave functions} of the total system, and has been called {\em canonical typicality}. 

We still need to answer when  we can assert  that the system is in equilibrium  and how do we  accurately model the very slow dynamical variables, especially if they form a continuum.
We also need to know  what happens when the system is at exactly $T=0$, where we can define a geometric entropy~\cite{Callan:1994} by partitioning a system into two parts $A$ and $B$ and define the von Neumann entropy (vNE):
\begin{equation}
S^{\mathrm{vN}}_{A}=-\mathrm{Tr}_{A}\rho_{A}\ln \rho_{A}=S_{B}^{\mathrm{vN}}=
-\mathrm{Tr}_{B}\rho_{B}\ln \rho_{B}.
\end{equation}
Here, the reduced density matrix $\rho_{A}$  is  obtained by tracing over the
degrees of freedom in $B$: $\rho_{A}=\mathrm{Tr}_{B}|\psi_{A B}\rangle\langle\psi_{AB}|$ and similarly for $\rho_{B}$.
In general, for a pure state $|\psi_{A B}\rangle$ of a composite system, which cannot be factored, the reduced density matrix is a mixture, and
the corresponding entropy is a good measure of entanglement. Therefoere vNE can play an important role for quantum phase transitions because correlations build up at the transition. It is also interesting to note that vNE is analogous in {\em form} to the classical Gibbs entropy
\begin{equation}
S_{G}= -k_{B}\int dX \rho(X) \ln \rho(X),
\end{equation}
where $\rho(X)dX$ is the phase space probability and the integral is over the phase space. If we choose $\rho(X)=1/W$, uniformly distributed over the available volume, $W$, of a macrostate, we arrive at the Boltzmann formula. Despite the similarity, vNE has a different meaning. It can give interesting information even at $T=0$ for two particles, namely the quantum mechanical entanglement of the particles. More curiously, we can give  a useful  physical meaning for even a single degree of freedom: for example, for a non-interacting particle in a random potential, we can  ask what is the amplitude of the complete wave function at a given site (a ket in the position basis), that is, how much does it participate at a given site. It can be finite for a localized state and zero for a delocalized state in the limit that the number of sites $N\to \infty$.

The role of multiplicity of dynamical scales is critical.
A simple well known example of two distinct time scales is the problem of ortho- and para-hydrogen.
The spins of the nuclei in a hydrogen molecule can be either in a triplet state or a singlet state.
The interaction between the nuclei is very small and so is the interaction between the nuclei 
and the electronic spins which are in a singlet state. Thus, the 
singlet-triplet conversion takes time, on the order of days, while the momenta of the 
molecules equilibrate on a microscopic time scale. In this example the number of nuclei in the
singlet state and the number of nuclei in the triplet state are separately constants of motion on
the time scale of a typical experiment. In considering the statistical mechanics
of this system one must take this fact into account, and add the free energies of these two subsytems
rather than adding the partition functions. Experimental observations strikingly confirm this fact. 

For widely separated scales, it is always clear what the relevant region of 
the phase space is. When this is not the case, and there is a
continuum of time scales, extending from very short microscopic scales to very long macroscopic
scales, this is difficult to determine. The common amorphous material, window glass, falls in this category. It is not  sufficient to know all the states and sum over all of
them; we must examine the actual dynamics of
the system.  However, the Boltzmann entropy formula is still usable. We could    determine the volume of the phase space sampled during  the observation time.  After all measurements are carried out on a single system not on ensembles. 

Many models of quantum phase transition consist of a single Hamiltonian, however many flavors of degrees of freedom we may have and however many coupling constants we may have. We are not restricted to a system weakly coupled to a reservoir of a generic character. A particular route to solving the problem may be to calculate the pure state (ground state) of the total system and the corresponding ground state energy. We might find that there is a sharp change of state, a phase transition,  at a given value of a coupling constant at which the ground state energy is non-analytic and the states on both sides of this transition are not unitarily related.  There are cases where this picture does not hold, however,  and we can have a perfectly analytic behavior of the ground state energy but a well defined non-analyticity of vNE characterizing distinctly  different correlations on the two sides of the transition. We do not know if this is a general principle signifying a quantum phase transition, but there cases where we can clearly demonstrate this phenomenon, as we shall see below. Somehow, Boltzmann's  notion of entropy continues to play an  important  role.

\section{Dissipation}

A convenient way to treat dissipative systems in quantum mechanics is to consider a system coupled to a reservoir and to integrate out the reservoir degrees of freedom. In this language dissipation refers to a one-way transfer of energy from a system to its environment, where it can be effectively lost~\cite{Weiss:1999}. If the energy levels of the environment is finely spaced,\footnote{It is convenient to use the terminology of  quantum mechanics.} there is little chance that energy that leaks out to the environment will ever return to the system in a physically meaningful time. For  quantum statistical mechanics,  as opposed to its classical cousin,  the notion of dissipation is relevant even in equilibrium, because statics and dynamics are intertwined through the complete set of energy eigenvalues of the total Hamiltonian, system plus the environment.  A quantum phase transition is certainly affected by dissipation, as opposed to a classical phase transition.

 If we represent a system plus its environment by a pure state wave function, there is no dissipation involved at zero temperature. Nonetheless, the internal correlations will still be affected  by what we would  call the environmental degrees of freedoms in an open system, except that we can no longer legitimately characterize them as due to dissipation.\footnote{We can still have a phase transition where the wave function undergoes a sharp change.With infinitely many degrees of freedom the two Hilbert spaces on the two sides of the transition will generically be unitarily inequivalent.} The discussion is not merely semantic, because for many  situations we can only measure the observables  of the system alone; the environment is important only to the extent that it influences the properties of the system. In such a coarse grained picture it is possible that the environment itself can be classified according to universality classes, and only the overall quantum numbers, such as charge, momentum, angular momentum etc. are relevant. 

Two crisp questions one can address are: (1) Given that we know that the classical system is dissipative and is well characterized by a few experimentally determined parameters, how does it behave when it is quantized? (2) Given that a quantum system in the ground state is well characterized by order parameters, how does the environment influence the statics and the dynamics of this order parameter and to what degree. I shall not address these questions further because much has already been said in answer to the first question by Feynman and Vernon~\cite{Feynman:1963}, Caldeira and Leggett~\cite{Caldeira:1983}, and Mohring and Smilansky~\cite{Mohring:1980}. With respect to the second question the pioneering work of John Hertz~\cite{Hertz:1976} has triggered a vast literature.

\section{Quantum Phase Transition}
Classical mechanics is fully described by Newton's laws, a set of differential equations for each degree of freedom. By following the solution, one should be able to predict every single outcome given a set of initial conditions. Yet, in the overwhelmingly majority of cases  this is not possible. The outcomes, the emergent states of matter, are strikingly unpredictable, especially when it comes to prediction of a phase transition. Phases of matter are strictly distinct and are not analytic continuations of each other. While Newton's laws are invariant under time-reversal and a given Hamiltonian may also be invariant under the same symmetry,  a phase may macroscopically violate this symmetry, a ferromagnet for example. In this respect quantum phases and  transitions between them are no different. They emerge  from quantum fluctuations of interacting infinitely many degrees of freedom.

\subsection{Infinite Number of Degrees of Freedom}

A mathematically precise solution of a model in two dimensions known as the Ising model by Onsager really clinches any possible doubts one may harbor. The same Hamiltonian, which is invariant under the flip of the Ising spins, can describe both a magnetized and an unmagnetized (disordered) state.~\footnote{Before Onsager's solution there was considerable doubt that this was possible---C. N. Yang, private communication} The response to an infinitesimal magnetic field below the phase transition is infinite, while above the transition the response to an infinitesimal field is infinitesimal. Even though both a magnetized and unmagnetised ensemble remain possible mathematical solutions, even a stray physical field can choose one state over the other. It would be  difficult to pretend that the actual physical state is a superposition of the two states related by spin flips, hence no magnetization. Experimental  observations bear out this conclusion to such a perfection that I can bet my life on it.

As is well known, the partition function of the Ising model is a sum 

\begin{equation}
Z= \sum_{\{s_{i}\}} \prod_{\langle i j \rangle}e^{-\beta S_{i}S_{j}},
\end{equation}
where the product is over nearest neighbor pairs and I have set the exchange constant to unity. The Ising spins $S_{i}$ take values $\pm 1$. As long as the number of spins, $N$, is finite, $Z $ is a polynomial of finite degree with real coefficients, and the free energy $F=-\frac{1}{\beta} \ln Z$ cannot possibly have any nonanalyticity at a real non-zero  temperature, hence no phase transitions. But once we entertain the possibility of $N=\infty$, the convergence of the sum is no longer guaranteed, and all bets are off. Now, of course, we are not truly interested in an infinite number of degrees of freedom. This would  not be physically meaningful. We are interested in  a sufficiently large number of degrees of freedom. How large is large? In a schematic language, given the experimental resolution $\epsilon$, we should choose a $\delta=1/N$ (or some power, but that is irrelevant), such that $\delta < \epsilon$.

The background of classical statistical mechanics and the role of an infinite number of degrees of freedom is a prelude in helping one understand quantum to classical crossover (transition), which is often ignored in discussions involving foundations of quantum mechanics. That quantum phase transition renders the system classical due to dissipation is an important piece of the puzzle. A genuine phase transition involves an infinite number of degrees of freedom. 

All we need is a single example where quantum to classical transition can be described in the language of a phase transition of an equivalent classical statistical mechanical problem. However, there is not just one but a whole class of problems where a $d$-dimensional quantum problem can be mapped onto a $(d+1)$-dimensional ($d$-refers to the spatial dimension and the remaining dimension is the imaginary time) classical statistical mechanical problem, and by analyzing  the classical problem, we can learn {\em some} important aspects of quantum mechanics. One of the simplest examples in this regard is a particle in a symmetric double-well potential. 

From standard undergraduate quantum mechanics, we know that the two lowest states are separated by a gap, $\Delta E$, related to the tunneling through the barrier, and a particle initially prepared in one of the wells will exhibit coherent oscillations, so characteristic of quantum mechanics. The same problem can be mapped onto a classical one-dimensional Ising model ($0\to 1$) from which we can infer the exact value of $\Delta E$, and all correlation functions in imaginary time and, by analytic continuation to real time, all the traditional aspects of quantum mechanics. We learn that as long as the barrier height is finite, the ground state is a linear superposition of two states separated by $\Delta E$ from the first excited state, which is an antisymmetric linear superposition. 

This classic example exemplifies all  the perplexing  aspects of quantum mechanics. Nonetheless, we can derive all its properties from the statistical mechanics of the classical  Ising model. Even more shocking is the problem  where this symmetric double well  is coupled to an infinite number of dissipative degrees of freedom with a special property termed as an Ohmic heat bath. This time the corresponding $(0+1)$-dimensional classical problem involves special long range interactions, which undergoes a phase transition as a function of the coupling to the environment in the limit that the number of Ising spins $N\to \infty$.  This translates into spontaneous symmetry breaking in the ground state of the quantum system~\cite{Chakravarty:1982}.  Parity is broken and one of the states will be selected by an infinitesimal asymmetric perturbation. The correlations within the broken symmetry state have effectively lost all vestige of quantum mechanics and have transformed a quantum  problem  to a classical problem. Here, then, is a problem that   is a tantalizing example of how quantum phase transition can illuminate foundations of quantum mechanics. 

\subsection{Broken Symmetry}
In this subsection the relevance of broken symmetry and quantum criticality will be discussed in the context of quantum measurement theory. The simplest case is that of a ferromagnet (say, the nearest-neighbor Heisenberg model to be more specific) in its ground state for a spin rotationally invariant Hamiltonian. For simplicity consider spin-$1/2$. The ground state of a ferromagnet does not respect the spin rotational symmetry symmetry of the Hamiltonian; an infinitesimally weak magnetic field can lock the ferromagnet in a given direction. Another way of expressing the same fact is to note that the two ground states which differ by an arbitrarily small angle are unitarily inequivalent in the limit of infinite number of spins. The theorems of Wigner, or Stone and von Neumann, regarding unitary equivalence  apply only to a finite number of degrees of freedom. 

\subsubsection{Unitary Inequivalence}

Let $|0\rangle$ to be the ground state  of a ferromagnet consisting of $N$ spins, which in the $S_{z}$ representation, is  
\begin{equation}
|0\rangle = |\uparrow_{1},\uparrow_{2}, \ldots \uparrow_{i}. \ldots \uparrow_{N}\rangle.
\end{equation}
Consider a new ground state where all spins are rotated by an angle $\theta$ in the $XZ$-plane be
\begin{equation}
|\theta\rangle = \exp\left[- i \frac{\theta}{2}\sum_{k=1}^{N}\sigma_{2}(k)\right]|0\rangle, \; 0< \theta \le \pi.
\end{equation}
The operators $\sigma_{\alpha}$ are the standard Pauli matrices.
It is trivial to show that the scalar product 
\begin{equation}
\langle 0|\theta\rangle = \left(\cos \frac{\theta}{2}\right)^{N} \to 0,\; \textrm{as} \; N\to \infty.
\end{equation}
We can build a Hilbert space by applying operators $\sigma_{-}(k)=\sigma_{1}(k)-i\sigma_{2}(k)$ on $|0\rangle$ and by Cauchy completion. Similarly, we can also start with
\begin{equation}
|\theta\rangle =U(\theta)|0\rangle
\end{equation}
and build up a Hilbert space based on the rotated Pauli matrices
\begin{equation}
\tau_{\alpha}(k)= U(\theta) \sigma_{\alpha}(k)U^{\dagger}(\theta)
\end{equation}
These Hilbert spaces are easily shown to be unitarily inequivalent, as any scalar product of a rotated state and the unrotated state vanishes in the limit $N\to \infty$.

\section{Measurement Theory}
This is a vast subject with a very long history. In this section we devote ourselves to a very specific neglected aspect, namely the role of broken symmetry and unitary inequivalence.
\subsection{Coleman-Hepp Model}

The above intuition regarding the unitary inequivalence of  two degenerate broken symmetry ground states of a ferromagnet can be used to build a measuring device. In a paper with the title,``Quantum Theory of Measurement and Macroscopic Observables'' K. Hepp~\cite{Hepp:1972} has discussed how probabilities are generated from probability amplitudes. In one of his models, the Coleman-Hepp model,  a particle with a kinetic energy, linear in momentum, zips along a chain consisting of spin-$1/2$ objects, flipping their spins. The linearity of the kinetic energy  leads to a wave packet that does not diffuse or change its form, which is a useful simplification.   Similarly, the model is further simplified by assigning zero energy to the bath of spins, so flipping a spin costs no energy whatsoever. These shortcomings can be easily repaired, but the  essential conclusions remain unchanged. Here we focus on  impementing the notion of unitary inequivalence discussed above in a pristine form.\footnote{In{\it Physics} Aristotle remarks  that we should begin by what is clear to us and then proceed to what is clear in itself.} It is of course understood that a macroscopic measuring device based on broken symmetry must have the necessary rigidity to be a valid measuring device and must be able to exchange energy with the system,  the particle in this case. Nonetheless, we shall see that the interference between the two wave packets of the particle can be destroyed by a dephasing mechanism (a form of decoherence) due to the  environment.

The Hamiltonian of the original Coleman-Hepp model is
\begin{equation}
H = v p +\sum_{i=1}^{N} V(x-x_{i}) \sigma_{1}(i),
\end{equation}
where $v$ is the velocity of the particle and $p=\frac{\hbar}{i}\frac{d}{dx}$ is the momentum operator conjugate to the coordinate $x$, while $x_{i}$ labels the lattice sites of the environment spins. The interaction $V(x)$ is supposed have a compact support vanishing beyond a range $r$, that is,
\begin{equation}
V(x) = 0 \; \textrm{for} \; |x| > r.
\end{equation}
It is useful to parametrize the strength of the interaction by a dimensionless quantity, which will later be interpreted as an angle of rotation of the bath spins:
\begin{equation}
\theta = \frac{1}{\hbar v}\int_{-\infty}^{\infty} dx V(x).
\end{equation}
Then the  the $S^{[N]}$-matrix is given by 
\begin{equation}
S^{[N]}= \lim_{t\to \infty, t'\to -\infty} U(t,t') =\prod_{k=1}^{N}S(k),
\end{equation}
where 
\begin{equation}
U(t,t')=e^{ivpt/\hbar}e^{-iH(t-t')/\hbar}e^{-ivpt/\hbar},
\end{equation}
and 
\begin{equation}
S(k)=\cos\theta-i\sigma_{1}(k)\sin\theta.
\end{equation}

Consider now a typical interference experiment where the initial state $|I\rangle$ is a direct product of a linear superposition of two wave packet states of the particle, $|\psi_{1}\rangle$ and $|\psi_{2}\rangle$ (normalized to unity), and the ground state of the environment $|0\rangle$,
\begin{equation}
|I\rangle = (|\psi_{1}\rangle+|\psi_{2}\rangle)|0\rangle.
\end{equation}
Imagine that only  $|\psi_{2}\rangle$ interacts with environment. Then the final state $|F\rangle$ is
\begin{equation}
|F\rangle = |\psi_{1}\rangle|0\rangle+S^{[N]} |\psi_{2}\rangle |0\rangle.
\end{equation}
The probability, $P$,  for the final state after the interaction with the environment  has taken place  is
\begin{eqnarray}
P&= &2+2 \Re\langle\psi_{1}|\psi_{2}\rangle \langle 0| S^{[N]}|0 \rangle,\nonumber \\
          &=&2+2 \left( \cos \frac{\theta}{2}\right)^N \Re\langle\psi_{1}|\psi_{2}\rangle \nonumber \\
          &\to & 2, \; N\to \infty.
\end{eqnarray}
Interference is therefore totally destroyed in the limit $N\to \infty$. The mechanism, as promised, is the unitary inequivalence of the rotated and unrotated Hilbert spaces in the infinite volume limit. Realistically, this inequivalence arises when  the ground state is a ferromagnet with a spontaneously broken symmetry.

The significance of the above result in the context of quantum measurement  problem has been disputed by J. S. Bell~\cite{Bell:1975}.
His criticisms are: (1) the limit $t\to \infty, \; t'\to -\infty$ ``never comes'', and (2) ``While for any given observable one can find a time for which the unwanted interference is as small as you like, for any given time one can find an observable for which it is as big as you do {\em not} like. ''  Bell is not entirely justified.  To quote from the abstract of Hepp's paper,
\begin{quotation}
... In several explicitly soluble models, the measurement leads to macroscopically different `pointer positions' and to a rigorous `reduction of the wave packet' with respect to all local observables.
\end{quotation}
The quotation marks are hints that neither the `pointer positions'  nor the `reduction of the wave packet' had a sense of finality in Hepp's elegant paper.  He further remarks
\begin{quotation}
For practical purposes it is not necessary to pass to infinite systems and times. However, one has to establish the existence of the limits $N\to \infty$ and $t\to \infty$ and the disjointedness of the resulting states of the system and apparatus, in order to be sure that in the finite approximations the error can be made arbitrarily small for sufficiently large $N$ and $t$. 
\end{quotation}
It is ironical that Bell  who made popular  the playful  acronym~\cite{Bell:1990} FAPP (for all practical purposes) seems not to appreciate the strength of Hepp's argument.
However, Bell does provide an example of a complex nonlocal observable for which the unwanted interference can be made as large as you do not like, and he  further remarks, 
\begin{quotation}
The continuing dispute about quantum measurement theory is not between people \ldots with different ideas about actual practicality of measuring arbitrarily complicated observables.
\end{quotation}
In my opinion, physical significance of an  arbitrarily complex non-local observable is debatable to say the least. Hepp's rigorous paper does have  an element of truth. He is more right than Bell. 

\subsection{Tunneling Versus Coherence}
There is a marvelous semiclassical analysis by Callan and Coleman that is worth revisiting~\cite{Coleman:1977}. The tunneling rate from a metastable well in a one-particle quantum mechanics is determined by the bounce formula for the tunneling rate
\begin{equation}
\Gamma =\hbar K e^{-S_{0}/\hbar},
\end{equation} 
where $K$ is the prefactor and $S_{0}$ the Euclidean action for a single bounce. It is easy to show that the bounces form a dilute gas with negligible interactions among them. When the particle in the metastable well is coupled to an Ohmic heat bath, Caldeira and Leggett solved the  bounce problem and showed that the quantum tunneling rate is reduced by friction~\cite{Caldeira:1983,Chang:1984}. They were also able to prove that bounces still form a dilute gas, again with negligible interaction.

For single particle quantum mechanics in a double well potential, the instanton (analog of bounce) calculation proceeds analogously and once again the dilute gas approximation works accurately. One discovers the familiar splitting between symmetric and antisymmetric states. This calculation can be easily generalized to periodic potential~\cite{Coleman:1977}. The instantons are much the same. The only novelty is  that when doing the dilute gas sum there is no constraint that instantons and anti-instantons must alternate. We sprinkle them freely along the real axis. 

\subsection{Quantum to Classical Transition}
\begin{figure}[h]
\begin{center}
\includegraphics[scale=0.5]{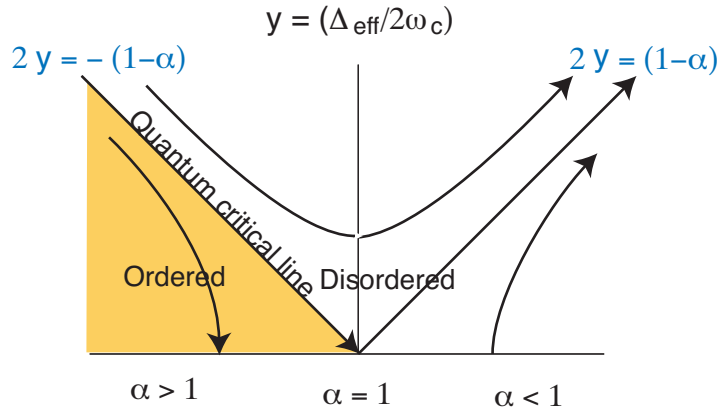}
\caption{The $T=0$ phase diagram of a two-level system, from Ref.~\cite{Kopp:2007}, coupled to an Ohmic heat bath and the corresponding renormalization group flows in the regime $|(1-\alpha)|\ll 1$ and $y\ll 1$. The renormalization group trajectories point along increasing values of the flow parameter $l=\ln(\tau \omega_c)$, where $\tau$ is imaginary time. The separatrix $2y=-(1-\alpha)$ separates a broken symmetry phase with finite magnetization from the disordered phase. At this critical line the magnetization is discontinuous even though the correlation time (imaginary) diverges.  At $\alpha=1/2$ the dynamics of the two-level system changes from an overdamped to an underdamped state, but there is no thermodynamic phase transition at this point.}
\end{center}
\label{fig:phase-diag}
\end{figure}

The situation changes dramatically in the double well problem, and {\em inter alia} the periodic potential problem, when we turn on Ohmic dissipation. The instantons now interact logarithmically, or in the equivalent description in terms of Ising spins the interaction is an inverse square interaction in imaginary time. The interaction between the instantons are now scale invariant and the dilute gas approximation fails dramatically, resulting in a phase transition as a function of the dissipation strength (cf. below) $\alpha \sim R_{Q}/R$ ($R_{Q}=h/4e^{2}$ being the quantum of resistance and $R$ the shunt resistance) and the fugacity of the instantons $y= \Delta_{\mathrm{eff}}/2\hbar\omega_{c}$, where $\hbar \omega_{c}$ is a high energy cutoff in the dissipative kernel, and $\Delta_{\mathrm{eff}}$ is the tunneling splitting ~\cite{Chakravarty:1982,Chakravarty:1985} sans instanton interactions but renormalized by dissipation, as shown in Fig.~\ref{fig:phase-diag} .

The continuous quantum phase transition for the  double well implies that the ground state is two-fold degenerate and the coherence is lost for sufficiently large dissipation due to the environment. This is none other than a quantum to classical transition, and as such it should valuable for a better understanding of the quantum measurement problem (or problems), as was the Coleman-Hepp model. The root of this broken symmetry is the orthogonality catastrophe~\cite{Anderson:1967} between the two degenerate states of the double well, which is $\sim (\omega/\omega_{c})^{\alpha}$, vanishing as $\omega\to 0$, where $1/\omega$ is the observation scale and $(1/\omega_{c})$ is a short time cutoff. Morally it is identical to the problem discussed in the context of the Coleman-Hepp model, and the role of orthogonality catastrophe was also recognized by Hepp in his original paper. It turns out that this orthogonality catastrophe can be cured by a renormalization group analysis~\cite{Chakravarty:1982} for $\alpha < 1$ for small $\Delta_{\mathrm{eff}}/\hbar \omega_{c}$, but not for $\alpha>1$.  A quantum phase transition separates the two regimes and the ensuing criticality has been extensively studied. In a loose sense one might say that an infrared divergent heat bath localizes the particle by repeated measurements (interactions) such that it looses its quantum coherent property. In other words, quantum interference is destroyed, as in the  Coleman-Hepp model.

What about a particle in a periodic potential coupled to Ohmic dissipation~\cite{Schmid:1983}? A myth has developed in our field that this problem is somehow different. The commonality of the double well and the periodic potential is often not recognized, although it was clearly recognized by Schmid himself~\cite{Schmid:1988}. As discussed above in the case of single particle quantum mechanics, the only novelty here is that  we can freely sprinkle instantons and anti-instantons without any constraint, but that they interact logarithmically leads again to a quantum phase transition with details that are a bit different.

\section{Von Neumann Entropy}
Entanglement is a unique feature of a quantum system and the von Neumann entropy (vNE)
 is a widely used measure of entanglement~\cite{Amico:2008}. 
 \subsection{A Warmup Exercise: Damped Harmonic Oscillator}
 Consider a single harmonic oscillator (momentum $p$, position $x$, mass $M$, and frequency $\omega_0$) coupled to an environment of harmonic oscillators. We examine the case of Ohmic dissipation, where the spectral function, $J(\omega)$, of the bath is defined by  the coupling constants $\{\lambda_n\}$, together with the masses $\{m_n\}$ and the frequencies $\{\omega_n\}$ of the oscillators comprising the bath. In this model $J(\omega)$ defined by  
\begin{equation}
J(\omega)=\frac{\pi}{2} \sum_n \frac{\lambda_n^2}{m_n \omega_n} \, \delta(\omega-\omega_n)
\label{spectraldens}
\end{equation}
is taken to be  $2 \pi \alpha \omega=\eta\omega$ for $\omega < \omega_{c}$ and zero for $\omega \geq \omega_{c}$.
The ground state expectation values of $x^2$ and $p^2$ are~ \cite{Weiss:1999}
\begin{eqnarray}
\langle x^2 \rangle & = & \frac{\hbar}{2M\omega_0} f(\kappa) \\
\langle p^2 \rangle & = & \frac{\hbar M \omega_0}{2}(1-2\kappa^2)f(\kappa)+\frac{2\hbar M \omega_0}{\pi} \, \kappa \ln(\frac{\omega_c}{\omega_0})
\end{eqnarray}
where $\kappa=\eta/2M \omega_0$ is the friction coefficient and 
\begin{equation}
f(\kappa)=\frac{1}{\pi \sqrt{\kappa^2-1}} \ln \left( \frac{\kappa+\sqrt{\kappa^2-1}}{\kappa-\sqrt{\kappa^2-1}} \right)
\end{equation}
At $\kappa=1$ the system crosses over from damped oscillatiory to overdamped behavior ($ \sqrt{\kappa^2-1}$ is to be replaced by $ i \sqrt{1-\kappa^2}$.  The function
$f(\kappa)$ is real for all $\kappa>0$ and has identical power series expansion regardless of the limits $\kappa \to1\pm 0$.

\begin{figure}[htb]
\begin{center}
\includegraphics[scale=0.5]{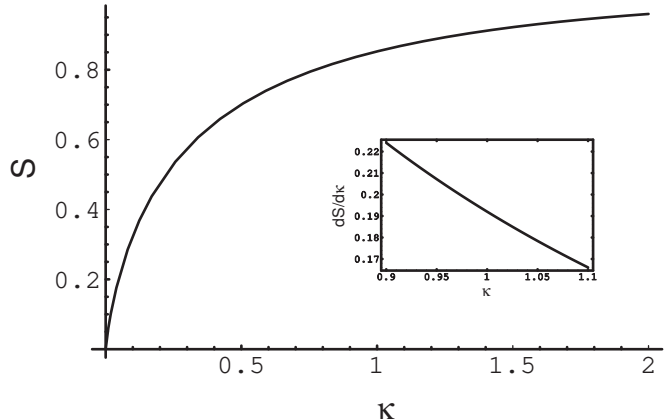}
\caption{Von Neumann entropy ($S$) of the damped harmonic oscillator, from Ref.~\cite{Kopp:2007}, with $\omega_c=100\, \omega_0$.  Both $S$ and its first derivative (inset) are perfectly smooth through the dynamic crossover at $\kappa=1$.  Higher-order derivatives are similarly well-behaved due to the analyticity of $f(\kappa)$ at $\kappa=1$.}
\end{center}
\label{fig:Sdho}
\end{figure}

At zero temperature, the normalized reduced density matrix for the damped harmonic oscillator has matrix elements \cite{Weiss:1999}
\begin{equation}
\langle x^{\prime} | \rho_A | x^{\prime\prime} \rangle = \sqrt{4b/\pi}\,e^{-a(x^{\prime}-x^{\prime\prime})^2-b(x^{\prime}+x^{\prime\prime})^2}
\end{equation}
where $a=\langle p^2 \rangle/2\hbar^2$ and $b=1/8\langle x^2 \rangle$.  To compute the von Neumann entropy, we first note \cite{Holzhey:1994}  that
\begin{equation}
\mathrm{Tr}(\rho_A \ln \rho_A)=\, \lim_{n\to1}\,\frac{\partial}{\partial n} \int dx^{\prime} \langle x^{\prime} | \rho_A^n | x^{\prime} \rangle
\end{equation}  
The subsequent mathematical manipulations described elsewhere~\cite{Kopp:2007}  result in the vNE of the damped harmonic oscillator, $S$,
\begin{equation}
S =-\frac{1}{2}\ln \left( \frac{4b}{a-b} \right) +\frac{1}{2} \sqrt{\frac{a}{b}}\,\ln \left( \frac{\sqrt{a}+\sqrt{b}}{\sqrt{a}-\sqrt{b}} \right)
\label{dhoentropy}
\end{equation}
Note that $S \to 0$ as $b\to a$, which corresponds to the minimum uncertainty $\sqrt{\langle x^{2}\rangle\langle p^{2}\rangle}=\hbar/2$. The uncertainty relation is satisfied only for $b\le a$. This  expression is analytic at $\kappa=1$  despite the transition from the overdamped  to the underdamped behavior because of the analyticity of $f(\kappa)$. The point $\kappa=1$ is {\em not} a point of quantum phase transition but a crossover in the dynamics. A plot of $S$ is shown in Fig.~\ref{fig:Sdho}.

\subsection{Double Well Coupled to a Dissipative Heat Bath}

We now return to the double well problem in the specific formulation known as the spin-boson model~\cite{Leggett:1987}  in which a two-level system is represented by a spin-1/2 degree of freedom and the bath is a collection of harmonic oscillators.  The Hamiltonian is
\begin{equation}
H_{\mathrm{sb}}=-\frac{1}{2} \, \Delta_{\mathrm{eff}} \, \sigma_1+ H_{\rm osc} + \frac{1}{2} \, \sigma_3\sum_n \lambda_n x_n .
\label{hamiltonian}
\end{equation}
The last term couples  $\sigma_{3}$ to the coordinates $\{x_n\}$ of the oscillators.  We shall consider an Ohmic heat bath, as defined above.

At zero temperature, this model has a quantum critical line separating a broken-symmetry phase with $\langle \sigma_3 \rangle =M_{0}\neq 0$ from a disordered phase with $\langle \sigma_3 \rangle = 0$ (see Fig.~\ref{fig:phase-diag})~ 
\cite{Chakravarty:1982}. The broken-symmetry state has an effective  classical description in which the environmental degrees of freedom are relaxed around a specific state of the qubit. The corresponding uncertainty is zero and so is the von Neumann entropy. In the disordered state, by contrast, an infinitesimal field has only an infinitesimal effect. Right at the quantum critical point the qubit is maximally entangled with the environment, as described below. So the phase transition can be aptly described as a classical-to-quantum transition and was treated by a renormalization group analysis~\cite{Chakravarty:1982}.

In the ground state of Eq.~\ref{hamiltonian}, the reduced density matrix of the spin degree of freedom is determined by the expectation values $\langle \sigma_1 \rangle$ and $\langle \sigma_3 \rangle$:
\begin{equation}
\rho_A=\frac{1}{2} \left( \begin{array}{cc}
                                             1+\langle \sigma_3 \rangle & \langle \sigma_1 \rangle \\
                                             \langle \sigma_1 \rangle & 1-\langle \sigma_3 \rangle
                                           \end{array} \right)
\end{equation}
After diagonalizing $\rho_A$, we can easily compute the ground state von Neumann entropy.
Hence the behavior of the entanglement at the transition follows directly from the behavior of $\langle \sigma_1 \rangle$ and $\langle \sigma_3 \rangle$.
Since the order parameter $\langle \sigma_3 \rangle$ is discontinuous at the transition, the von Neumann entropy also jumps by an amount $\Delta S$.  It can be shown~\cite{Kopp:2007} that  the magnitude of this jump is, to leading order in $y$,
\begin{equation}
\Delta S =  \ln 2+(y/2) \ln y
\end{equation}
In the limit of vanishing $y$, the system goes from being unentangled ($\langle \sigma_1 \rangle=0$, $\langle \sigma_3 \rangle =1$) to being maximally entangled ($\langle \sigma_1 \rangle=\langle \sigma_3 \rangle =0$) as it enters the disordered state.  Note that this result depends crucially on a proper treatment of the broken symmetry---without the jump in the order parameter, $S$ would be continuous through the transition.  The von Neumann entropy is discontinuous (with a singular derivative) even though the correlation length in imaginary time diverges at the transition.  Because this divergence takes the form of an essential singularity, it does not leave a strong signature in other quantities at the critical point.

In addition, the spin-boson model also undergoes a dynamic crossover at $\alpha=1/2$, from damped oscillations to an overdamped decay.  The von Neumann entropy should be analytic at $\alpha=1/2$ because no phase transition occurs at this point, similar to the damped harmonic oscillator problem above.  This appears to be consistent with the calculations in Ref.~\cite{Kopp2:2007}

\subsection{Disordered Systems}

We shall give two examples~\cite{Kopp:2007,Jia:2008} of disordered systems involving non-interacting electons, probably many more exist.
For the Anderson localization transition in three
dimensions (3D) and the integer quantum Hall (IQH) plateau
transition in two dimensions  the ground state energy does
not exhibit any non-analyticity~\cite{Edwards:1971}. In contrast, vNE exhibits 
non-analyticity that can be determined from multifractal scaling~\cite{Evers:2008}.
It should be emphasized, however,  that because of the
single particle and disorder-dominated nature of these quantum
phase transitions, entanglement as characterized by vNE and its
critical scaling behavior are fundamentally different from
those calculated for interacting pure systems. 

In a non-interacting electronic system close to a disordered
critical point, the wave function intensity at energy $E$,
$|\psi_E(r)|^2$ fluctuates strongly at each spatial point $r$
and exhibits a multiplicity of fractal dimensions. This non
self-averaging nature of the wave function intensity can be expressed 
by the generalized inverse participation ratios
$P_q$ obeying the finite size scaling
\begin{align} \label{definition of Pq}
P_q(E) \equiv \sum_{r} \overline{ \left| \psi_E(r)\right|^{2q}}
\sim L^{-\tau_q} \, \mathcal{F}_q\big[(E-E_C)L^{1/
\nu}\big].
\end{align}
Here $L$ is the system size, $\nu$ is the localization length exponent given by  $\xi_E
\sim |E-E_C|^{-\nu}$, and $\tau_q$ is the multifractal
spectrum.  The overbar denotes average over distinct
disorder realizations. $\mathcal{F}_q(x)$ is a scaling function
with $\mathcal{F}_q(x\rightarrow 0) = 1$ close to the critical
point $E=E_C$. When $E$ is tuned away from $E_C$, the system
either tends towards an ideal metallic state with $P_q(E) \sim
L^{-d(q-1)}$ ($d$ being the number of spatial dimensions) or
becomes localized with $P_q(E)$ independent of $L$.

There is no obvious way to define entanglement in the language of particles. Nonetheless, surely vNE can be defined, and in fact 
entanglement can also be defined, but in the second-quantized Fock space. ~\cite{Zanardi:2002}
A simple analysis shows that the vNE is 
\begin{align}\label{single site entropy sum}
S(E)  &= - \sum_{r \in L^d} \Bigl[ |\psi_E(r)|^2   \ln
|\psi_E(r)|^2 \nonumber \\ & \quad + \left(1- |\psi_E(r)|^2
\right) \ln \left( 1- |\psi_E(r)|^2 \right)\Bigr].
\end{align}

The second term inside the square bracket in
Eq. \eqref{single site entropy sum} can be dropped, since
$\left| \psi_{E}(r)\right|^2  \ll 1$ at all $r$ when the
states are close to the critical energy. 
The disorder averaged (denoted by overbar) entropy using
multifractal scaling in Eq. \eqref{definition of Pq} is
\begin{align}\label{EE summed over all sites}
\overline{S}(E)\approx -\frac{dP_{q}}{dq}\bigg|_{q=1} \approx
\frac{d\tau_q}{dq}\bigg|_{q=1} \ln L - \frac{\partial
\mathcal{F}_q}{\partial q} \bigg|_{q=1}.
\end{align}
The general form of the scaling function
$\mathcal{F}_q$ is not known but we can get the approximate $L$ dependence
 in certain limiting cases. At exact criticality
when $\mathcal{F}_q \equiv 1$ for all 
$q$, we get
\begin{align}\label{EE critical scaling}
\overline{S}(E) \sim \alpha_1 \ln L,
\end{align}
where  $\alpha_1 ={d\tau_{q}/dq}|_{q=1}$
is unique for each universality class. From the discussion
following Eq. \eqref{definition of Pq}, the leading scaling
behavior of $\overline{S}(E)$ in the ideal metallic and
localized states is given by $d \ln L$ and $\alpha_1 \ln \xi_E$
respectively. From the limiting cases, we see that in general,
$\overline{S}(E)$ has the leading scaling form~\cite{Jia:2008}
\begin{align}\label{approximate form for EE of single energy state}
\overline{S}(E) \sim \mathcal{K}[(E-E_C)L^{1/\nu}] \ln L,
\end{align}
where the coefficient function $\mathcal{K}(x)$  decreases from
$d$ in the metallic state to $\alpha_1$ at criticality and then
drops to zero for the localized state. 

\subsubsection{Anderson Localization}

Consider the three dimensional 
Anderson model,~\cite{Abrahams:1979} on a cubic lattice. The Hamiltonian is
\begin{equation}\label{Hamiltonian_Anderson}
    H=\sum_i V_i c_i^\dag c_i-t\sum_{\langle i,j\rangle}(c_i^\dag
    c_j+h.c)
\end{equation}
where $c_i^{\dag}$($c_i$) is the fermionic creation
(annihilation) operator at the site $i$ of the lattice, and the
second sum is over all nearest neighbors. We set $t=1$ and
$V_i$ are random variables uniformly distributed in the range
$[-W/2,W/2]$.  It is known that  extended states appear at the
band center for $W$ less than  $W_c=16.3$,
and  the\ localization
length exponent is~\cite{Slevin:2001} $\nu=1.57\pm 0.03$.

The analysis leading to Eq. \eqref{approximate form for EE of
single energy state} also holds  at
a single energy, say $E=0$, as disorder is tuned through $W_c$; the states at $E=0$ evolve from fully
metallic to critical and then to localized behavior. The leading scaling form  of the
vNE is 
\begin{align}\label{generalscaling2}
\overline{S}(E=0,w,L) \sim \mathcal{C}(wL^{1/\nu})\ln L,
\end{align}
where  $w=(W-W_c)/W_c$  and $\mathcal{C}(x)$ is a scaling function. As mentioned above, 
$\mathcal{C}(x) \to d$ as $w
\to -1$, $\mathcal{C}(x) \to 0$ as $w \to \infty$, and
$\mathcal{C}(x) =\alpha_1$ when $w=0$.
 We can also average Eq.
\eqref{EE summed over all sites} over the  entire  band of
energy eigenvalues and construct the vNE,
\begin{align}
\overline{S}(w,L) = \frac{1}{L^3}\sum_{E} \overline{S}(E,w,L),
\end{align}
where $L^3$ is also the total number of states in the band.
Then using Eq. \eqref{approximate form for EE of single energy
state} and Eq. \eqref{generalscaling2}, one can show that close
to $w=0$,
\begin{align}\label{generalscaling1}
\overline{S}(w,L) \sim C
+L^{-1/\nu}f_{\pm}\big(wL^{1/\nu}\big)\ln L
\end{align}
where $C$ is an \emph{L independent} constant, and $f_\pm(x)$
are two universal functions corresponding to the regimes
$w>0$ and $w<0$.

\begin{figure}[htb]
\label{Scaling_Anderson1}
\begin{center}
\includegraphics[width=\linewidth]{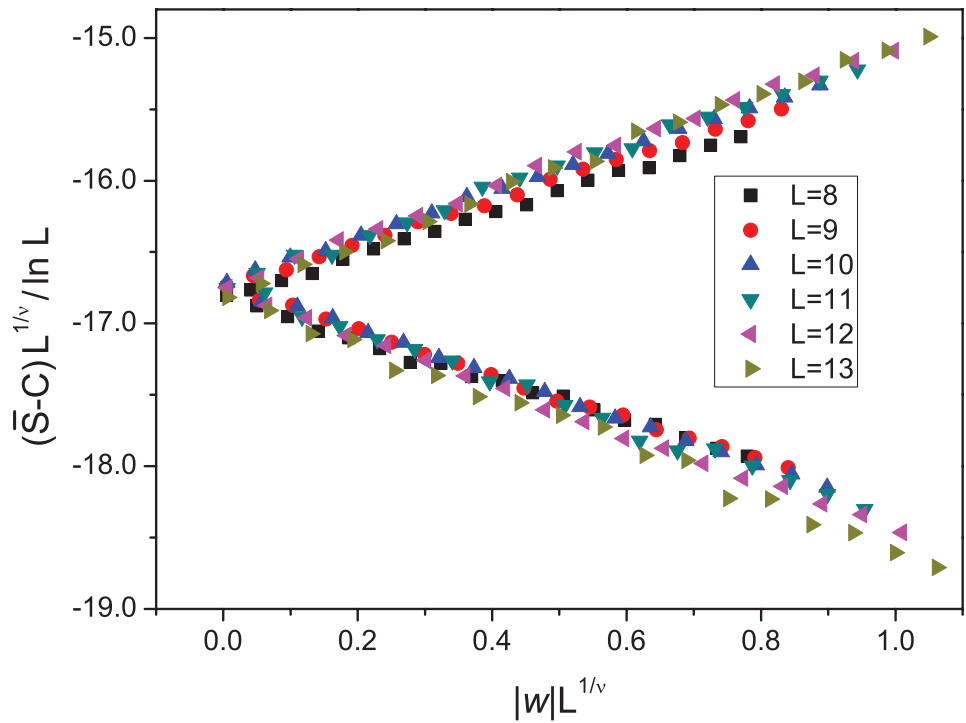}
\caption{Scaling curve in the 3D Anderson model. With the
    choice of $\nu=1.57$ and $C=12.96$, all data collapse to 
    universal functions $f_\pm(x)$. The two branches correspond
    to $w<0$ and $w>0$, from Ref.~\cite{Jia:2008}.}
    \end{center}
\end{figure}

Numerical results for  $\overline{S}(w,L)$ is  shown in
Figure~\ref{Scaling_Anderson1}. The data collapse is performed  with the choicef $\nu=1.57$, and the nonuniversal
constant $C=12.96$ is determined by a powerful algorithm described elsewhere~\cite{Kopp:2007}. The success of
data collapse clearly reflects the non-analyticity of the von Neumann entropy and the validity of the multifractal analysis.

\subsubsection{Integer Quantum Hall Plateau Transitions}

The Hamiltonian for the integer quantum Hall problem in a magnetic field $B$ can be
expressed in terms of the matrix elements of the states $|n,k\rangle$, where  $n$ is the  Landau
level index, and  $k$ is the wave vector in the $y$ direction. Let us focus  on the lowest Landau level $n=0$, with the
impurity distribution $\overline{V(\mathbf{r})V(\mathbf{r'})}=V_0^2\delta(\mathbf{r}-\mathbf{r'})$; the elements
of the random matrix $\langle 0,k|V|0,k'\rangle$ can be generated as in Ref.~\cite{Huckestein:1995}.

Consider a  square with linear dimension $L=\sqrt{2\pi}
Ml_B$, where $l_B=(\hbar/eB)^{1/2}$ is the magnetic length and
$M$ is an integer with periodic boundary conditions imposed in
both directions. Upon discretization with a mesh of size
$\sqrt{\pi} l_B/\sqrt{2}M$ the eigenstates 
 $\{|\psi_a\rangle=\sum_k
\alpha_{k,a}|0,k\rangle\}_{a=1}^{M^2}$  and the corresponding eigenvalues $\{E_a\}_{a=1}^{M^2}$ are obtained. The energies
are measured relative to the center of the lowest Landau band in units of $\Gamma=2V_0/\sqrt{2\pi}l_B$, and for 
 each eigenstate the wave function in real space
is
\begin{equation}
    \psi_a(x,y)=\langle x,y|\psi_a\rangle=\sum_k
    \alpha_{k,a}\psi_{0,k}(x,y) \label{wave function_IQHE}
\end{equation}
where $\psi_{0,k}(x,y)$ is the lowest Landau level wave
function.Unfortunately, the dimension of the  Hamiltonian matrix , $M^{2}\times
M^2$, can be very large. Therefore,  we
compute only those states $|\psi_a\rangle$ whose energies lie
in a small window $\Delta$ around a preset value of $E$, i.e.
$E_a\in[E-\Delta/2,E+\Delta/2]$. We ensure that $\Delta$ is
sufficiently small ($0.01$),  containing at the same time enough states iwithin the interval (at least 100
eigenstates).

\begin{figure}[htb]
\label{scaling_IQHE}
\begin{center}
\includegraphics[width=\linewidth]{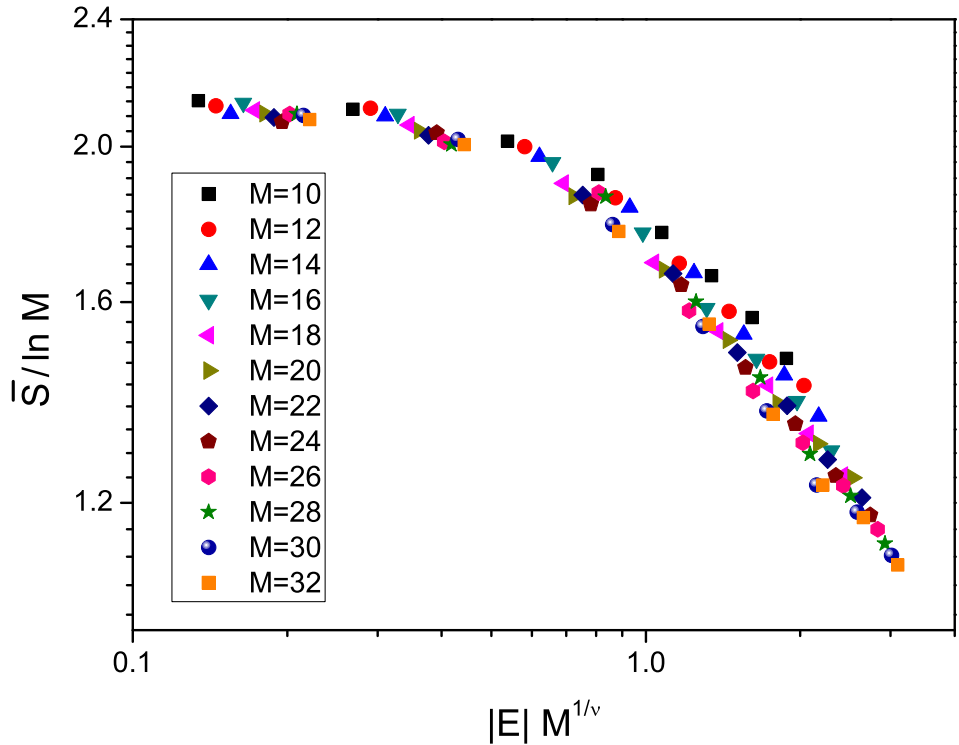}
\caption{Scaling of the von Neumann entropy  $\overline{S}(E)$ for integer quantum Hall effect. $M$ instead of $L$ is used in the data
collapse with the accepted value of $\nu=2.33$, from Ref.~\cite{Jia:2008}.}
\end{center}
\end{figure}

The next step involves  breaking up the $L\times L$ square into
nonoverlapping squares $\mathcal{A}_i$ of size $l\times l$,
where $l=l_B\sqrt{\pi/2}$ independent of the system size $L$.
We now  compute the coarse grained probability
$\int_{(x,y)\in\mathcal{A}_i}|\psi_a(x,y)|^2\mathrm{d}x\mathrm{d}y$.
The computation of the vNE  follows the
procedure described above.
Finally, by averaging over states in the interval $\Delta$, the
vNE $\overline{S}(E, L)$ is obtained at the preset energy $E$.
The scaling form of $\overline{S}(E,L)$ given by Eq.
\eqref{approximate form for EE of single energy state} with
$E_C = 0$ is
$\overline{S}(E,L)=\mathcal{K}(|E|L^{1/\nu})\ln L$. Good
agreement with the numerical simulations is seen in
Fig.~\ref{scaling_IQHE}~\cite{Jia:2008}. Note that there is only one branch, as states at all energies 
except that at the center of the band are localized. The plateau to plateau transition takes 
place at the band center, and the non-analyticity of this phase transition  is correctly captured by vNE.

\subsubsection{Infinite Randomness Fixed Point}

Recently an intriguing set of results have been obtained for one-dimensional quantum criticality for strong disorder and have uncovered what is known as infinite randomness fixed point.~\cite{Fisher:1995} There is some work in higher dimensional systems, but much less. The principal defining feature of infinite randomness fixed point is its scaling properties. In conventional quantum critical points, energy and length satisfy a dynamic scaling relationship defined by $E\sim L^{-z}$, where $z$ is the dynamic scaling exponent. In contrast, for infinite randomness fixed point the excitation energy behaves as 
\begin{equation}
E\sim e^{-L^{\psi}},
\end{equation}{
where $\psi$ is a universal exponent. The low energy distribution of  couplings, $\rho(J)$,  obey an unusual distribution,
\begin{equation}
\rho(J) \sim \frac{1}{J^{1-\chi/\Gamma}},
\end{equation}
where $\chi$ is a universal constant and $\Gamma= \ln \omega_{c}/\omega$, where $\omega$ is the measurement energy scale and $\omega_{c}$ is the largest bare coupling.

One-dimensional infinite randomness fixed points are random analogs of pure $(1+1)$-dimensional conformal field theories whose von Neumann entropies satisfy~\cite{Holzhey:1994} 
\begin{equation}
\lim_{N\to \infty} S = \frac{c}{3} \log_{2} N
\end{equation}
This entanglement entropy is defined by partitioning the system into $A$ and its complement $B$, where $A$ is a connected set of $N$ sites in an otherwise infinite system. The parameter $c$ is the universal conformal charge defining the theory. $S$ is bounded away from the critical point as $N\to \infty$. It is quite remarkable that an equivalent classical theory determines the information entropy of a critical quantum system in a universal manner. Thus it is even more remarkable that the von Neumann entropy corresponding to infinite randomness fixed points satisfy an analogous universal relation~\cite{refael:2009},
\begin{equation}
\lim_{N\to \infty} S = \frac{c_{\mathrm {eff}}}{3} \log_{2} N
\end{equation}
One would have intuitively thought this not to be possible because quenched randomness brings along with it its own probability measure. In either case note the violation of the area law, implying a length scale corresponding to the $AB$ boundary over which entanglement decays. Entanglement becomes infinitely long ranged as the correlations become infinitely long ranged at a quantum critical point. The values of $c_{\mathrm {eff}}$ have been calculated in a number of cases. One believes that these are exact results because the real space renormalization group methods in one dimension become asymptotically exact for for strong disorder. Table~\ref{tb:ccharge} summarizes some interesting comparison of pure and random systems (taken from Ref.~\cite{refael:2009}).

\begin{table}[htb]
\caption{\; Conformal charges of pure and random systems, from Ref.~\cite{refael:2009}}
\vskip 0.5cm
\begin{tabular}{ccc}
Model &$c$ & $c_{\mathrm{eff}}$\\\hline \\
$XXZ$-Heisenberg & $1$ & $\ln 2$ \\ \\
Transverse field Ising & $1/2$ & $1/2 \ln 2$\\ \\
Spin-1 Heisenberg (breakdown of Haldane phse) &$3/2$ & $1.232$ \\ \\
$SU(2)_{3}$ (Fibonacci anyons) & $4/5$ & $0.702$\\ \\
\end{tabular}
\label{tb:ccharge}
\end{table}

\section{Disorder and First Order Quantum Phase Transition}

The effect of disorder on continuous classical phase transitions has been studied over many decades, but less is known about its effect on  first order  transitions. Imry and Wortis~\cite{Imry:1979} argued that arbitrarily weak disorder can actually round a classical first order transition. Subsequently, Hui and Berker~\cite{Hui:1989} and Aizenman and Wehr~\cite{Aizenman:1990} have made important contributions to this topic.    

Two important questions are: (a) Can disorder convert  a first order quantum phase transition  to a continuous one? (b) If the answer to this question is yes, what are the universality classes, if any? 
We shall assume that  disorder couples to the Hamiltonian in such a way that its symmetry is unchanged,  for example,  to  nearest  neighbor bonds, or more generally  to energy-like variables. 
A site random field  on the other hand breaks symmetry explicitly.  Let a  tuning parameter $g$ control the  relative magnitudes of two noncommuting terms in the Hamiltonian,  resulting in a first order transition. 

We would like to present an argument~\cite{Goswami:2008} that the coexistence of phases is not possible at this transition because of disorder, and the quantum fluctuations do not have a scale. If this is true, and if the state corresponding to $g=0$ is still a broken symmetry state (this is why we imposed the specific requirement on disorder earlier) and the $g=\infty$ is a quantum disordered state, the conclusion must be that the transition has been converted to a quantum critical point.

The proof is by contradiction.  Assume that there is coexistence of phases at the  first order transition at $g_{c}$. However, in the presence of disorder there will be local fluctuations of  $g_{c}$.   Thus, within a putative quantum disordered region, randomness can nucleate an ordered region of linear dimension $L$, with a gain in the volume energy $\propto L^{d/2}$ (assuming central limit theorem), while the price in the surface energy is $\propto L^{d-1}$ (assuming discrete symmetry). The same is true for a putative ordered region. Therefore, for $d  <  2$ (discrete) , the picture is that of a ``domain within domain'', and there is no scale.  In contrast, for $g < g_{c}$ nucleation of one broken symmetry phase within another does not gain any energy (disorder does not break the relevant symmetry), but the surface energy is increased. Therefore, by contradiction, coexistence of phases is not possible, and the transition at $g_{c}$ must  be continuous. We have tacitly assumed that the transition involves a broken symmetry. If this is not the case, there is no particular reason for a sharp transition to remain at $g_{c}$, and the disorder will simply smear out the transition.

Continuous symmetry leads to a subtlety. While one may be tempted  to argue that the domain wall energy is $L^{d-2}$~, as in the Imry-Ma argument~\cite{Imry:1975}, this is generally incorrect. If at the domain wall   the amplitude of the order parameter vanishes, the domain wall energy is still $L^{d-1}$, as for discrete symmetry and the previous result holds. However, if the transition  is driven by tuning a ``magnetic field'' that changes the state from one broken symmetry direction to another, the domain wall energy is indeed $L^{d-2}$, and the borderline dimensionality is $d=4$. Since Mermin-Wagner theorem dictates that there is no long range order in $d=2$ at any finite temperature, regardless of the order of the transition, for the classical case the question is moot at $d=2$.

There are no simple arguments known to us for the borderline dimensionalities,  but from the rigorous version of the Imry-Ma argument for the random field case, it is safe to conjecture that the above argument should also hold for these cases because of a close connection between the two problems noted by Imry~\cite{Imry:1984}. Note that the dynamic critical exponent $z$ does not enter the above argument --- all we need is the extensivity of the ground state energy and its normal fluctuations in the thermodynamic limit. The principal disordering agent that washes out the coexistence is the fluctuation due to impurities and not quantum fluctuations. Quantum fluctuations can only help the process of smoothing the coexistence. Of course the fate of the system in dimensions higher than 2  must depend on the quantum fluctuations. In addition, the actual dynamics of the system must involve these fluctuations as well. The Harris criterion that determines the influence of disorder at a critical point, {\em inter alia} a quantum critical point,  does depend on the  exponent $z$. A nice way to see this is to rephrase the Harris criterion~\cite{Harris:1974} in terms of an argument by Mott~\cite{Mott:1981}. On one hand, disorder in a domain of linear dimension, $\xi$, the correlation length of the pure system,  will give rise to fluctuations of the quantum critical point $g_{c}$ of fractional width $\Delta g \sim \xi^{-d/2}$. On the other hand, $\Delta g$ must be less than the reduced distance from the quantum criticality implied by $\xi$, that is $\sim \xi^{-1/\nu_{d+z}}$, for the criticality to remain unchanged. Hence, $\nu_{d+z}>2/d$. Otherwise, the system may be described by a new disorder fixed point for which the same relation will apply with the replacement of the critical exponent of the pure system by the critical exponent of the new fixed point, as in the theorem of Chayes {\em et al}~\cite{Chayes:1986}. In either case $z$ appears  because the relevant length scale close to the critical point is the {\em diverging correlation length}, $\xi$.  By contrast, the argument for rounding of a first order transition is restricted by a {\em finite correlation length}, hence the balance is between the volume energy and the surface energy of a fluctuating domain nucleated by disorder. 

In order to substantiate our argument we studied~\cite{Goswami:2008},  using both a perturbative renormalization group and a real space decimation procedure, the one-dimensional quantum  random $N$-color Ashkin-Teller model in the regime  in which the pure model has a first order quantum phase transition. The corresponding classical problem in two spatial dimensions have renormalization group flows that curl back to the pure decoupled Ising fixed point at least for weak coupling~\cite{Cardy:1999}. In the quantum case the flows are drastically different and are towards the strong coupling  regime. It is therefore not possible to reach a definitive conclusion. On the other hand the strong coupling real space decimation technique shows that for a range of parameters depending on $N$ the flow is to the infinite randomness fixed point. No firm conclusions could be drawn beyond this regime. Recently, rigorous mathematical analysis has been brought to bear on this problem by Greenblatt, Aizenman and Lebowitz~\cite{Greenblatt:2009}. This proof of rounding of first order quantum phase transition is different from the classical proof~\cite{Aizenman:1990}. Clearly, more work is necessary to fully elucidate this interesting problem, with possibly far reaching consequences.

\section{Outlook}
One of the simplest quantum phase transitions is the dissipative transition in a double well potential in the presence of  Ohmic dissipation, which belongs to the same universality class as the transition in the Kondo problem, as the coupling is tuned from ferro to antiferromagnetic, as well as to the inverse square Ising model in one dimension~\cite{Chakravarty:1995,Voelker:1998}. As of today, there is no direct experimental evidence of this phenomenon, which is a pity. The prediction~\cite{Chakravarty:1984} for $\alpha > 1$  of the finite temperature incoherent tunneling rate between the wells seems to have some support from experiments~\cite{Han:1991}. This rate is 
\begin{equation}
\frac{1}{\tau} = \frac{\sqrt{\pi}\Delta_{\mathrm{eff}}(\alpha)^{2}}{2\omega_{c}}
\frac{\Gamma(\alpha)}{\Gamma(\alpha+1/2)}\left[\frac{\pi k_{B}T}{\hbar\omega_{c}}\right]^{2\alpha-1}.
\end{equation}}
Instead, the attention has shifted more to reducing dissipation as much as possible in various Josephson devices to observe coherent oscillations of the double well~\cite{Friedman:2000,Chiorescu:2003}. This is considered to be a miniature prototype for Schr\"odinger's cat. It can be debated as to the extent to which the phase difference of a Josephson device or equivalently the flux variable in a superconducting interference device can be considered a ``macroscopic variable'', which appears to be  a semantic issue. These are indeed collective degrees of freedom, but the energetics are determined by ``microscopic'' scales. Very little experimental attention has been paid to explore the quantum critical dynamics of how one of the basic models of quantum mechanics is influenced by dissipation. As mentioned above, the influence of dissipation in quantum tunneling from a metastable state does not count, because no quantum criticality is involved. Given that convincing experimental demonstrations of quantum criticality are so few and far between and generally so complex due to complex material issues that it seems hopeless to make further progress without a clear cut study of the simplest possible example. In any case, as I have argued this is a well studied example of a quantum to classical transition (FAPP) whose mathematics is firmly grounded at this time.

The second overarching theme of this article has been how disorder influences quantum phase transitions. While much is understood for classical phase transitions, very few results are available for transitions at zero temperature. This is again a pity because there is considerable theoretical depth to this problem. In fact, practical applications abound as well. Is it possible that many experimental sightings of quantum critical points are in reality first order transitions in disguise, rounded by disorder? If so, how does it enrich our understanding? An interesting question is the uses of von Neumann entropy. There appear to be some quantum phase transitions that exhibit no non-analyticity of the ground state energy and yet their existence can be hardly denied, as they signify transition between two distinct states of matter. In this respect, it was quite remarkable that von Neumann entropy should exhibit the requisite non-analyticity. Perhaps, the fundamental criterion should be replaced. So far we have only found such examples in disordered systems involving non-interacting electrons, Anderson localization in three dimensions and plateau-to-plateau transition in integer quantum Hall systems. Are there more, especially involving interacting systems? Is there a theorem? I conjecture that such results are only possible in a disordered systems where the fundamental driving mechanism is fluctuations driven by disorder that belong to a different class from quantum fluctuations triggered by a tuning coupling constant. It will be interesting to tackle the fractional quantum Hall problem from the perspective of the Jain construction~\cite{Jain:2007}, as this maps the problem to an essentially non-interacting problem.

Another problem involving quantum criticality in  disordered systems is the infinite randomness fixed point. A number of important theoretical calculations involving entanglement entropy has shown that the renormalization group flow is to new fixed points, different from the pure system fixed points, but with universal amplitudes of logarithmic entanglement entropies. Do these fixed points reflect the same properties of conformal invariance of pure systems? Or, are the mathematical underpinnings
different? Are there higher dimensional problems that can be solved in a similar manner?

Finally, a more pressing issue is the role of dissipative phase transition in a number of important fields of current research, to name a few,  phase slips in nanowires,~\cite{Bezryadin:2000,Lau:2001,Bollinger:2006,Goswami:2006,Refael:2007,Bollinger:2008,Bezryadin:2008} $c$-axis conductivity~\cite{Chakravarty:1994} and local quantum criticality~\cite{Chakravarty:1988} in high temperature superconductors~\cite{Aji:2007}.  There is also much interest in low-temperature properties of very thin superconducting wires. The key process of interest  are  quantum phase slips, a virtual depletion of the superconducting density that allows the  system to tunnel to a different value of the supercurrent. The rate of this process depends not only on the ÒbareÓ fugacity of the phase slips, defined by the rate of an individual tunneling event, but also importantly on the interaction between individual quantum phase slips. The backbone of the theoretical work is the logarithmic interaction between the quantum phase slips, which serves almost as a paradigm to whole class of similar problems~\cite{Mooij:2006,Khlebnikov:2008,Meidan:2007}. Interesting results have been obtained~\cite{Khlebnikov:2008}: it has been argued that for a short wire there is no distinction between a superconductor and an insulator. Even an insulator can support a weak Josephson current. Nonetheless, there is a range of parameters for which a short nanowire can act as an insulator down to unobservable low temperatures.
\section*{Acknowledgments}
I acknowledge a grant from the National Science Foundation (Grant No. DMR-0705092) and discussions with David Garcia-Aldea,  Joel Lebowitz, Chetan Nayak,  Gil Refael and David Schwab.

\end{document}